\documentstyle[sprocl,epsfig]{article}

\def\be{\begin{equation}}
\def\ee{\end{equation}}
\def\bea{\begin{eqnarray}}
\def\eea{\end{eqnarray}}

\begin{document}

\title{A NEW UNIFIED EVOLUTION EQUATION}
\author{Jyh-Liong Lim\footnote{This work is collaborated with
Hsiang-nan Li, Department of Physics, National Cheng-Kung University,
Tainan, Taiwan}}
\address{Department of Electrophysics, National Chiao-Tung University, \\
Hsinchu, Taiwan\\E-mail: u8221805@cc.nctu.edu.tw}
\maketitle\abstracts{We propose
a new unified evolution equation for parton distribution
functions appropriate for both large and small {\em Bjorken
$x$}. Compared with the Ciafaloni-Catani-Fiorani-Marchesini
equation, the cancellation of sofe poles between
virtual and real gluon emissions is made explicit without introducing
infrared cutoffs, and next-to-leading contributions to the Sudakov 
resummation can be included systemmatically in this equation.}

\section{Introductions}
  As far as we know, the conventional evolution equations covering different 
kinematic regimes will be different equations. Such as: The {\bf DGLAP} ({\em 
Dokshitzer-Gribov-Lipatov-Altarelli-Parisi}) equation governs the large
logarithmic corrections due to high momentum transfer in the intermediate
{\em Bjorken x} region. The {\bf BFKL} ({\em Balitsky
-Fadin-Kuraev-Lipatov}) equation deals with the large logarithmic
corrections coming from the small {\em Bjorken x} region. The
{\bf CCFM} {\em Ciafaloni-Catani-Fiorani-Marchesini}) equation, as a 
unified approach of these above two, is appropriate for both intermediate
and small {\em x} regime. But the derivation of
these evolution equations based on the idea that the leading
logarithm contribution can be produced by summing the ladder
diagrams one by one to all orders. Also, by locating the momentum flows
of the rungs of the radiated gluon ladder diagrams at different orderings
, different evolution equations was obtained.\cite{AP,BFKL,CCFM}

Recently, we develop the resummation idea of
{\bf Collins-Soper-Sterman} which was
used to resum the double logarithmic correction of the parton
distribution function in hard QCD scattering process to the 
conventional evolution equations. We can easily reproduce all these 
above equations basing on the resummation idea instead of the complex
ladder diagram calculation. Also, we can introduce modified
evolution equation(Modified BFKL equation) to consider extra
kinematic variable dependence.\cite{L4} In 
this short paper, a new unified 
evolution equation innovated by resummation idea is proposed to
unify the kinematic regime that the DGLAP and BFKL work seperately.

\section{From Resummation Idea To A New Unified Evolution
equation}

\subsection{Resummation Idea} 
\label{subsec:resum}

  We study the unintegrated gluon distribution function in the axial gauge,
\bea
F(x,k_T,p^+)&=&\frac{1}{p^+}\int\frac{dy^-}{2\pi}\int\frac{d^2y_T}{4\pi}
e^{-i(xp^+y^--{\bf k}_T\cdot {\bf y}_T)}
\nonumber \\
& &\times\frac{1}{2}\sum_\sigma
\langle p,\sigma| F^+_\mu(y^-,y_T)F^{\mu+}(0)|p,\sigma\rangle\;,
\label{deg}
\eea
where $|p,\sigma\rangle$ denotes the incoming hadron with light-like
momentum $p^\mu=p^+\delta^{\mu +}$ and spin $\sigma$, and $F^+_\mu$ is the
field tensor.

Because of the scale invariance of $F$
in the gauge vector $n$,
$F$ must depend on $p^+$ via the ratio $(p\cdot n)^2/n^2$. Hence,
we have the chain rule relating $p^+d/dp^+$ to $d/dn$ :
\bea
p^+\frac{d}{dp^+}F=-\frac{n^2}{v\cdot n}v_\alpha\frac{d}{dn_\alpha}F\;,
\label{cph}
\eea
with $v_\alpha=\delta_{\alpha +}$ a dimensionless vector along $p$. The
operator $d/dn_\alpha$ applies to a gluon propagator, 
then the differentiated gluon attaches to a special vertex,
which is read off 
from the Eqs.~(\ref{cph}),\cite{CS}
\bea
{\hat v}_\alpha=\frac{n^2v_\alpha}{v\cdot nn\cdot l}\;.
\label{va}
\eea

Employing the Ward identities, it 
reduces to our master equation of resummation idea
with the special vertex at the outmost end.\cite{CS,L1} 
As shown in Fig.~1(a), we obtain: 
\be
p^+\frac{d}{dp^+}F(x,k_T,p^+)=2{\bar F}(x,k_T,p^+)\;,
\label{cc}
\ee

\begin{figure}[hbt]
\begin{tabular}{c@{\qquad}c}
\mbox{\epsfig{file=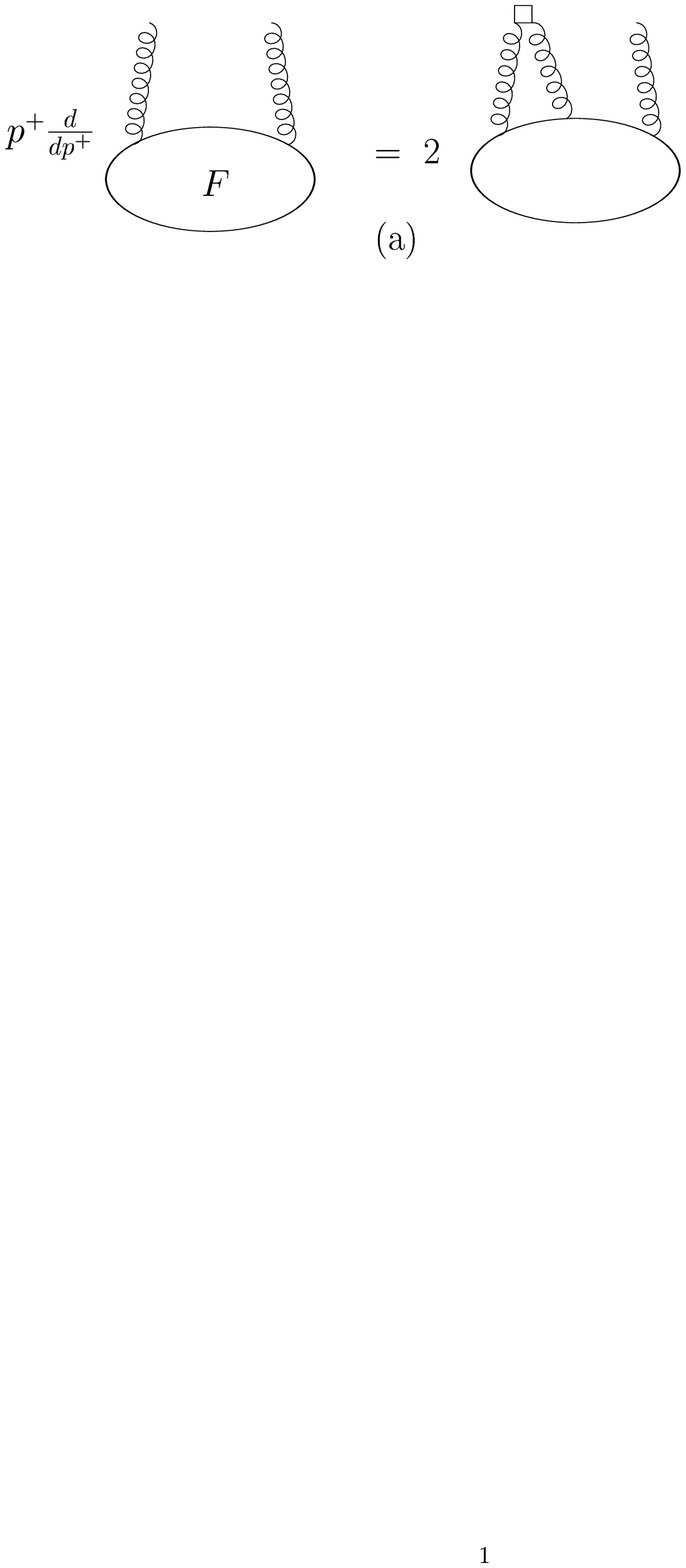,width=3.0cm}}\hfill
\mbox{\epsfig{file=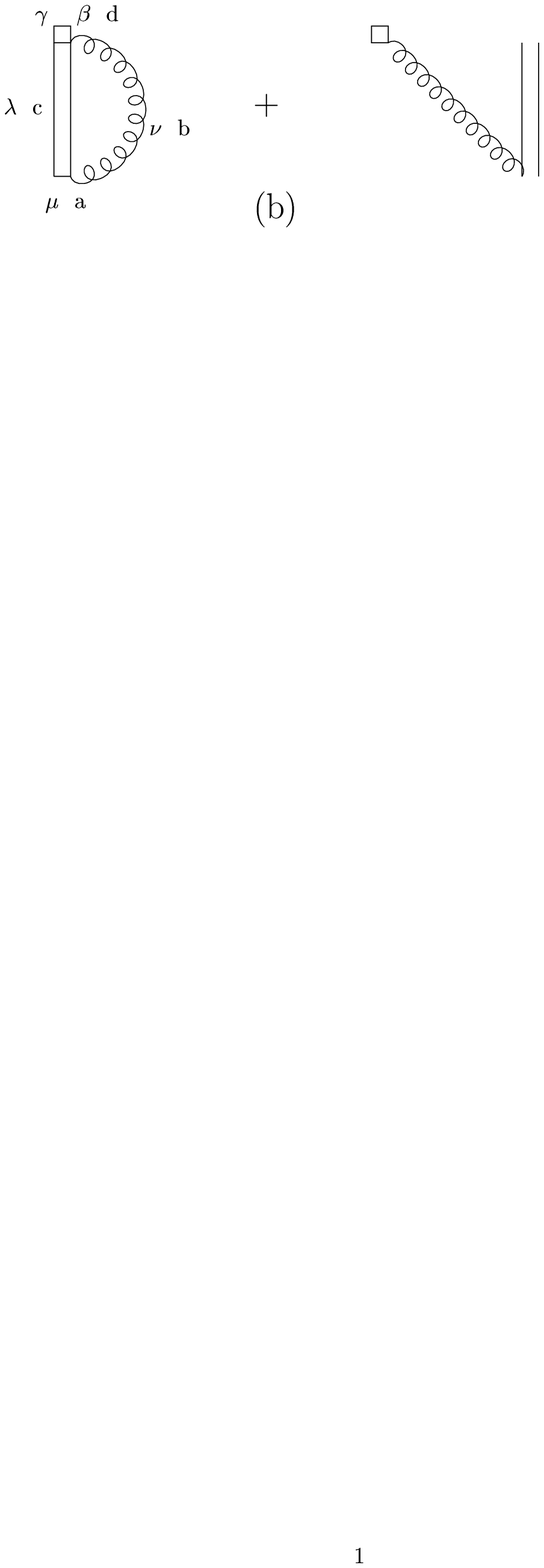,width=4.0cm}}\hfill
\mbox{\epsfig{file=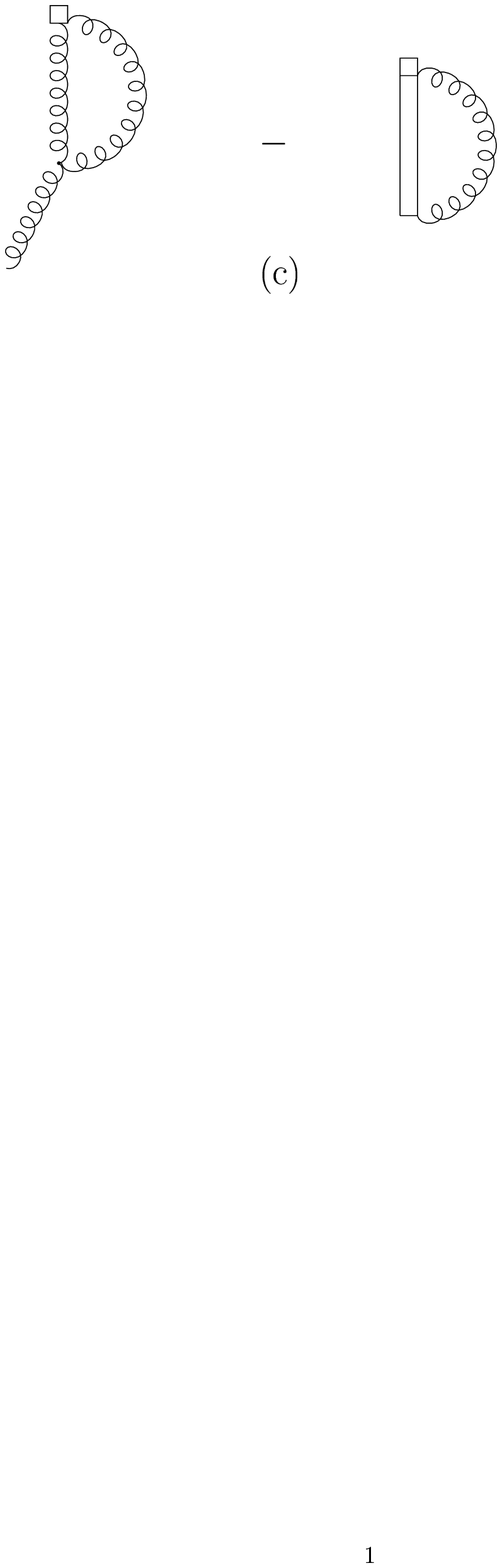,width=4.5cm}}
\end{tabular}
\vfill
\caption{ (a) The derivative $p^+dF/dp^+$ in the axial gauge. (b) The
lowest-order subdiagrams for $K$. (c) The lowest-order subdiagrams for $G$.}
\end{figure}

\subsection{A New Unified Evolution Equation}
First, we Fourier transform Eq.~(\ref{cc}) into the $b$ space, 
$b$ being the conjugate variable of $k_T$. 
The leading regions of the loop momentum flowing through the special
vertex are soft and hard, so
we factorize subdiagrams
containing the special vertex order by order 
(Figs.~1(b) and 1(c)). 

For the contribution from Fig.~1(b), We reexpress the function $F$ as
\bea
& &F(x+l^+/p^+,b,k^+)=\theta((1-x)p^+-l^+)F(x,b,k^+)
\nonumber\\
& &\hspace{1.5cm} +[F(x+l^+/p^+,b,k^+)-\theta((1-x)p^+-l^+)F(x,b,k^+)]\;.
\label{fre}
\eea
  The contribution from the first term 
of Eq.~(\ref{fre}) is combined with virtual gluon correction, leading to
\bea
K(x,b\mu,\alpha_s(\mu))&=&-{\bar \alpha}_s(\mu)\left[
K_0(2(1-x)p^+\nu b)+\ln \frac{b\mu}{2}+\gamma_E\right]\;,
\eea
the other from the
second term in Eq.~(\ref{fre}) is denoted as:
\bea
{\bar F}'_{s}
&\approx &{\bar\alpha}_s(k^+)\int_x^1 \frac{d\xi}{\xi}
\frac{F(\xi,b,k^+)-F(x,b,k^+)}{1-x/\xi}\;.
\label{sf2}
\eea

While Fig.~1(c) gives  
\begin{eqnarray}
G(k^+/\mu,\alpha_s(\mu))&=&-{\bar \alpha}_s(\mu)\left[
\ln\frac{2k^+}{\mu}+\ln\nu\right]\;,
\end{eqnarray}

However, the ultraviolet divergence of $K$ and $G$
cancel each other, the sum of $K$ and $G$ will be
RG invariant. Therefore, the RG solution will be
\bea
& &K(x,b\mu,\alpha_s(\mu))+G(k^+/\mu,\alpha_s(\mu))=
\nonumber \\
& &\hspace{1.0 cm}K(x,1,\alpha_s(k^+))+G(1,\alpha_s(k^+))-s(x,b,k^+)\;,
\label{shkg}
\eea
with
\bea
& &K(x,1,\alpha_s(k^+))={\bar \alpha}_s(k^+)\left[\ln(1-x)+\ln(p^+b)+
\ln 2\nu\right]\;,
\label{ka}\\
& &G(1,\alpha_s(k^+))=-{\bar \alpha}_s(k^+)\ln 2\nu\;,
\\
& &s(x,b,k^+)=\int_{1/b}^{k^+}\frac{d{\bar\mu}}{\bar\mu}
\left[\gamma_K(\alpha_s({\bar\mu}))
+\beta\frac{\partial}{\partial g}K(x,1,\alpha_s({\bar\mu}))\right]\;,
\label{se}
\eea

Absorbing the term ${\bar \alpha}_s\ln(1-x)F(x,b,k^+)$ 
, we arrive at the plus distribution $1/(1-x/\xi)_+$
as splitting function. Our evolution equation becomes
\bea
p^+\frac{d}{dp^+}F(x,b,k^+)&=&-2\left[s(b,k^+)
-{\bar \alpha}_s(k^+)\ln(p^+b)\right]F(x,b,k^+)
\nonumber\\
& &+2{\bar\alpha}_s(k^+)\int_x^1 \frac{dz}{z}
\frac{F(x/z,b,k^+)}{(1-z)_+}\;,
\label{ue1}
\eea


By including the Sudakov form factor and spiltting function,
our new unified evolution equation, as
a revised version of the CCFM equation, is very convicing.

\section{Conclusion}
In nature, the master equation of the Collins-Soper-Sterman 
resummation idea is equivalent to the evolution equation of
the gluon distribution function. After carefully analysing
the infrared and ultraviolet poles of evolution kernel, a
new unified evolution equation can be proposed for
both large and small $x$. The infrared finiteness is explicit
in both the Sudakov resummation and the splitting function. The 
next-to-leading logarithmic information are also included in the
Sudakov form factor systematically.

\section*{Acknowledgements}
This work is supported by National Science Council of R.O.C. under the
Grant No. NSC87-2112-M-006-018.

\vfill
\section*{References}

\end{document}